\documentclass[prx,twocolumn,floatix,amssymb,showpacs,amsmath,superscriptaddress]{revtex4-1}
\pdfoutput=1
\usepackage{graphicx}
\usepackage{epsfig}
\usepackage{amsmath,amssymb}
\usepackage{ dsfont }
\usepackage{color}

\begin{document}

\newcommand{\ket}[1]{\ensuremath{\left|{#1}\right\rangle}}
\newcommand{\bra}[1]{\ensuremath{\left\langle{#1}\right|}}
\newcommand{\quadr}[1]{\ensuremath{{\not}{#1}}}
\newcommand{\quadrd}[0]{\ensuremath{{\not}{\partial}}}
\newcommand{\slpar}{\partial\!\!\!/}
\newcommand{\gtrescero}{\gamma_{(3)}^0}
\newcommand{\gtresuno}{\gamma_{(3)}^1}
\newcommand{\gtresi}{\gamma_{(3)}^i}

\title{ Efficient universal quantum channel simulation in IBM's cloud quantum computer}

\author{Shi-Jie Wei}
\thanks{These authors contributed equally to this work.}
\address{ State Key Laboratory of Low-Dimensional Quantum Physics and Department of Physics, Tsinghua University, Beijing 100084, China}

\author{Tao Xin}
\thanks{These authors contributed equally to this work.}
\address{ State Key Laboratory of Low-Dimensional Quantum Physics and Department of Physics, Tsinghua University, Beijing 100084, China}

\author{Gui-Lu Long}
\email[Correspondence and requests for materials should be addressed to G.L.L.: ]{gllong@tsinghua.edu.cn}
\address{ State Key Laboratory of Low-Dimensional Quantum Physics and Department of Physics, Tsinghua University, Beijing 100084, China}
\address{ Tsinghua National Laboratory for Information Science and
Technology,  Beijing 100084, P. R. China.}
\address{ Collaborative Innovation Center of Quantum Matter, Beijing 100084, China}

\date{\today}

\begin{abstract}
The study of quantum channels is the fundamental field and promises wide range of  applications, because any physical process can be represented as a quantum channel transforming an initial state into a final state.
Inspired by the method performing non-unitary operator by the linear combination of unitary operations,
we proposed a quantum algorithm  for the simulation of universal single-qubit channel, described by a convex combination of  'quasiextreme' channels corresponding to four Kraus operators, and is scalable to arbitrary higher dimension.  We demonstrate the whole algorithm experimentally using the universal IBM cloud quantum computer and study properties of different qubit quantum channels.  We illustrate the quantum  capacity of  the general qubit quantum channels, which quantifies the amount of quantum information that can be protected. The behaviour of quantum capacity  in different  channels  reveal which types of noise processes can support information transmission, and which types are too destructive to protect information. 
There is   a general agreement between the theoretical predictions and the experiments, which strongly supported our method.  By realizing  arbitrary qubit channel, this work provides a universal way to explore various properties of quantum channel and novel prospect of quantum communication.
 \end{abstract}


\maketitle

\section{Introduction}
Since Feynman \cite{Fey82}proposed the idea of  quantum computer and  envisioned  the possibility of efficiently simulating quantum systems,  significant progress has been made in closed system quantum simulation.  Quantum simulation can   efficiently simulate   the dynamics of diverse systems\cite{Fey82,Lloyd} in condensed matter \cite{ap1,ap2} , quantum chemistry\cite{ap3}  , and high-energy physics\cite{ap4,ap5,ap6},  which is intractable on classical computers.  Moreover,  every  practical quantum system is open system because of the inevitable coupling to the  environment. Thus, quantum simulation of open system is an equally important and more general subject to explore. However, open quantum system simulation  is still in the early stages of development and  is concentrated  on
simulating Markovian dynamics by Lindblad master equation \cite{master1,master2,master3, master4,master5,master6}, which remains largely unexplored.
The quantum simulation of open system  promises powerful applications in a class of physical problem, such as preparing various special state \cite{state1,state2,state3, state4,state5},  thermalizing in  spin-boson systems and complex many fermion-boson  systems\cite{therm1,therm2}, studying nonequilibrium dynamics\cite{therm3}. 
Contrary to common sense, dissipative dynamics which is not necessarily unitary can be utilized to  perform universal quantum computation\cite{DD}.

Given the importance of the simulation of open quantum system,
efficiently performing quantum channels which represent the most general quantum dynamics possible is critical.  A straightforward way suggested by the Stinespring dilation theorem  \cite{Stine} for the simulation of open quantum systems is to  enlarge the system to include the environment,  which can be regarded as a bigger closed quantum system. Then, we can perform Hamiltonian-generated unitary transformation as same as in the closed system,  which means we can implement a channel as a unitary operator on an expanded Hilbert space. 
The evolution of the  density matrix~\cite{N}
 \begin{align}\label{q1}
  \rho^{'} =tr_{env}(U(\rho \otimes \rho_{env})U^{\dagger}),
 \end{align}
where $ \rho^{'} $ is the density matrix of the final state of principal system , $tr_{env}$ is a partial trace over environment and $ U $ is time evolution operator imposed on the  total system.  The disadvantage   in this method is that  the expanded Hilbert space dimension is at most  dimension $n^3$ of the original system because of an environment of dimension $n^2$ is necessary, which make it inefficient on high dimension. In recent years, many works have been done for achieving  channels in special cases \cite{his1,his2,his3,his4,his5,his6},  or generating arbitrary channels with significant  failure probability\cite{his7}.  It is worth to recall the brilliant idea of Wang $et~al$ which realised any qubit channel by Kraus operators with single qubit gates and controlled NOT gates \cite{prlw}, makes it possible to implement with current technology.

Here we present a new method which can realize universal qubit channels  deterministically with controlled NOT operations. In contrast with the method suggested in \cite{prlw}, the approach is a total quantum algorithm without a  classical random number in  generator and realise four Kraus operators  simultaneously. This algorithm requires  two qubits maximum as ancillary system to simulate the environment by performing the controlled  operations  on the single-qubit work system. Moreover, there are only single direction controlled  operations  from ancillary system to work qubit which are  not dependent on the state of the  single-qubit work system,  making the method more general and scalable in higher dimension with ancillary quantum resource in log$_2(d)$  qubits order. We realize the universal single-qubit channel  corresponding to four Kraus operators  \cite{math} simultaneously and can obtain the density information of work qubit under  any single Kraus operator. The algorithm is performed in IBM's quantum cloud computer,  and the behaviour of entanglement fidelity, entropy  and coherent information of different qubit quantum channels which play fundamental role in characterizing  the channel  capacity are explored.
We numerically  calculate the capacity of all qubit channels and analyse the behaviour of three important types quantum channels capacity which are general noises  affecting the quantum system. The calculation gives a metric on  how reliable  and efficient  of a quantum system to process information undergoes  a special channel. 

\section{Quantum algorithm to realise universal quantum channel }

Mathematically,  a completely positive and trace preserving linear map is a quantum channel, denoted as the set $\mathfrak{P}(S\mapsto S')$,  connecting system $S$ to system $S'$.  In particular,  if there exist unitary channels ${\cal U}^A \in \mathfrak{P}(S\mapsto S)$ and ${\cal U}^B \in \mathfrak{P}(S' \mapsto S')$  satisfy
\begin{eqnarray}
\Phi = {\cal U}{^B} \circ \Lambda \circ {\cal U}^A \;,\label{UNIEQ}
\end{eqnarray}
two maps $\Phi$, $\Lambda \in \mathfrak{P}(S\mapsto S')$  are unitarily equivalent\cite{RMP}.

A CPTP map expression is also equivalent to  operator sum (or Kraus) representation\cite{JA,Choi,N},
 \begin{equation}
 \large \Phi (\rho) = \sum_{j = 1}^r K_j \rho K_j^\dagger.
\end{equation}
where $\{K_j\}$ are Kraus operators on ${\cal H}_s$ which is the Hilbert space of open system and satisfy the completeness  conditions $\sum_j K_j^\dag K_j =I$. 
The Kraus rank $r = \mathrm{rank}(\tau) \leq d^2$ ($d$ is the dimension of ${\cal H}_s$)  is the number of non-zero  Kraus operators  guaranteeing that a Kraus representation exists with no more than $d^2$ elements.  In the case  of  single qubit channel,  we need at most four Kraus operators  to construct a Kraus representation. Specifically, by defining the operator $K_j$ as
${_E\langle} j|
U |0\rangle_E$ of ${\cal H}_S$, with  $\{ |j\rangle_E\}$
is an orthonormal basis of environment $E$, Kraus representations and the Stinespring  dilation are mutual correspondence\cite{RMP}. 

Qubit channels  are  CPTP transformations $\Phi \in \mathfrak{P}(S\mapsto S)$
that map the initial states  into final states in the same  two dimensional quantum system, denoted $\Phi: \mathcal{M}_2 \rightarrow \mathcal{M}_2$. A linear map $\Phi $ on $\mathcal{M}_2$ can also be represented by a unique $4 \times 4$ matrix ${\bf T}$ with 12 independent parameters, which is easy to characterize qubit channels\cite{math}. 
\begin{eqnarray}
{\bf T} =  \left(
 \begin{array} {cc}
     1 & {\bf 0} \\  
     {\bf t} & {\rm T}
      \end{array} \right)
     \end{eqnarray} 
where $\rm T$ is a $3 \times 3$ matrix , ${\bf t}$
is column vector and satisfy ${\bf T}$ is real. Corresponding to the Bloch ball representation which is more geometrical, this map transforms  the state ball into an ellipsoid expressed as

\begin{equation}
\Phi (\frac{1}{2} I+ \frac{1}{2} \text{\bf r} \cdot \text{\bf s})=\frac{1}{2} I+ \frac{1}{2} (\text{\bf t}+\text{\bf r}) \cdot \text{\bf s}
\label{eq:Trep.equiv}
\end{equation}
where  ${\sf s} = (\sigma_x, \sigma_y, \sigma_z)^{\top}$ is a column vector and $ \sf{r} = (r_x, r_y, r_z)\in  \mathbb{R}^3$ of length  $|\sf{r}|\leqslant 1$. 

For any CPTP map, there always exists an equivalent relationship between two maps. It is  
$\label{eq:KRpolarS}
\Phi(\rho) = {\cal U} ^B\Big[{\Phi}_{ {\bf \Lambda}} \Big]{\cal U}^{A} ,$
where  $ {\Phi}_{ {\bf \Lambda}}$ is a diagonal form via the singular-value decomposition\cite{math2}. $ {\Phi}_{ {\bf \Lambda}}$ is in  the closure of the extreme points corresponding to a $4 \times 4$  parameterization matrix ${\bf T}$   satisfying
\begin{equation}
{\bf T} = \left( \begin{array} {cccc}
 1 & 0 & 0 & 0 \\  0 & \cos u & 0 & 0 \\
   0 &  0 & \cos v & 0 \\
\sin u \sin v &  0 & 0 & \cos u \cos v
\end{array} \right)
\end{equation}
where $u \in [0,2\pi)$ and $v \in [0,2\pi)$. It is straightforward to prove that  this trigonometric parameterization map $ {\Phi}_{ {\bf \Lambda}}$   can be obtained by the
Kraus operators
\begin{equation}
	K_0=\begin{pmatrix}\cos\beta &0\\0&\cos\alpha\end{pmatrix}, \quad
	K_1=\begin{pmatrix}0&\sin\alpha\\\sin\beta&0\end{pmatrix},
\end{equation}
where $\alpha=(\mu+\nu)/2$ and $\beta=(\mu-\nu)/2$. According to the theorem in~\cite{math},  any stochastic map on  $\mathcal{M}_2$ can be written as a convex combination of two maps $ {\Phi}_{ {\bf \Lambda}}$ in the closure of the extreme points.  Namely, an arbitrary single-qubit channel  can be realized via four Kraus operators
\begin{eqnarray}
K_0=\sqrt{P}\begin{pmatrix}\cos\beta_{1} &0\\0&\cos\alpha_{1}\end{pmatrix}\\\nonumber
K_1=\sqrt{P}\begin{pmatrix}0&\sin\alpha_{1}\\\sin\beta_{1}&0\end{pmatrix}\\\nonumber
K_2=\sqrt{1-P}\begin{pmatrix}\cos\beta_{2} &0\\0&\cos\alpha_{2}\end{pmatrix}\\\nonumber
K_3=\sqrt{1-P}\begin{pmatrix}0&\sin\alpha_{2}\\\sin\beta_{2}&0\end{pmatrix}.
\label{kkkk}
\end{eqnarray}
where $ P $ is the probability from $ 0 $ to $ 1 $.

\begin{figure}
\centering
\includegraphics[width=0.48\textwidth]{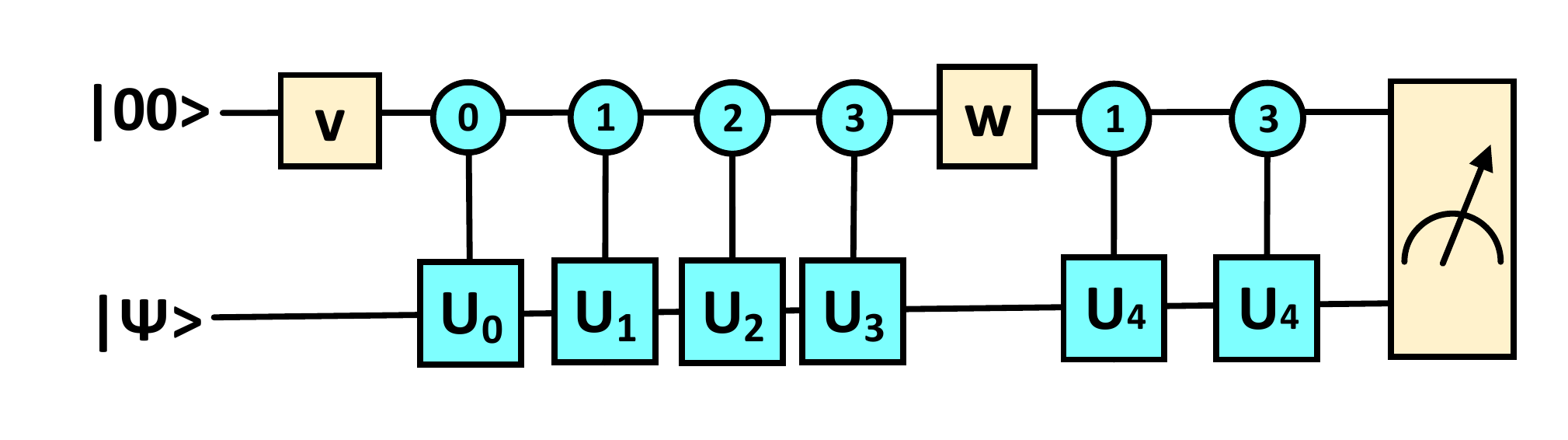}
\caption{  Quantum circuit  to implement the convex combination of two quasiextreme channels.  $|\Psi\rangle$ denotes the initial state of principal system, while environment is prepared in the $|00\rangle$ state. $V, W$  are unitary operators to product superposition  state and combine the controlled unitary operation respectively.  The squares represent unitary operations and the circles represent the state of the controlling qubit. Unitary operations $U_{0}$,  $U_{1}$, $U_{2}$ and  $U_{3}$  are activated only when the auxiliary qubit is $|00\rangle$ ,  $|01\rangle$ , $|10\rangle$ and  $|11\rangle$, respectively. $U_{4}$ is active when  the auxiliary qubit is $|01\rangle$ ,  or  $|11\rangle$.  } \label{faa}
\end{figure}

Considering $  K_{j}$ is a bounded linear operator in a finite dimensional Hilbert space which can be decomposed  into a sum of unitary operators, such that we adopt the duality quantum computing \cite{r1,r2,r3,r4,r5,r6,r7,r8,r9,QIP,SC} to perform the arbitrary single-qubit channel. In duality quantum computing, the work system with initial state $|\Psi\rangle$ and the $d$-dimension ancillary system with initial state $|0\rangle$  are coupled together. The corresponding quantum circuit of the algorithm is further shown in Fig. \ref{faa}.

In the following, the detailed parameters $V, W$ and the controlled gates $U_i \otimes |i\rangle \langle i|(i=1,...,d-1)$ are determined for efficiently simulating the universal single-qubit quantum channel illustrated in equation (\ref{kkkk}).   To make sure $W$ and $V$ being unitary, the Kraus operators are rewritten as
\begin{eqnarray}
K_{0}&=&\sqrt{P} \big[(\cos \beta_{1} + \cos \alpha_{1})\frac{I}{2} \,+\,
 (\cos \beta_{1} - \cos \alpha _{1})\frac{Z}{2}  \,  \big]\\
\nonumber  
K_{1}&=&\sqrt{P} \big[(\sin \beta_{1} + \sin \alpha_{1})\frac{X}{2}   \,+\,
 (\sin \beta_{1} - \sin \alpha_{1})\frac{ZX}{2}  \, \big]  \\   
\nonumber     
K_{2}&=&\sqrt{1-P}\big[(\cos \beta_{2} + \cos \alpha_{2})\frac{I}{2}    \,+
\, (\cos \beta_{2} - \cos \alpha_{2} )\frac{Z}{2}  \,\big]  \\
\nonumber       
K_{3}&=&\sqrt{1-P} \big[(\sin \beta_{2} + \sin \alpha_{2})\frac{Z}{2}   \,+
\, (\sin \beta_{2} - \sin \alpha_{2})\frac{ZX}{2} \,\big]   
\end{eqnarray}\label{kk}
where $ I $ is identity matrix and $ Z $, $ X $ are  pauli matrix. We define unitary operators $U_{0}=U_{2}=I$, $ U_{1}=U_{3}=Z$, $ U_{4}=X$.

The unitary operator   $V$ is
 \begin{eqnarray}
V&=&\left(                 
  \begin{array}{cccc}   
  \sqrt{\frac{P(1+\cos(\beta_{1}-\alpha_{1}))}{2}}  & N & N & N\\  
   \sqrt{\frac{P(1-\cos(\beta_{1}-\alpha_{1}))}{2}}  & N & N & N\\  
 \sqrt{\frac{(1-P)(1+\cos(\beta_{2}-\alpha_{2}))}{2}}  & N & N & N\\
  \sqrt{\frac{(1-P)(1-\cos(\beta_{2}-\alpha_{2}))}{2}} & N & N & N\\
  \end{array}
\right)  \\\nonumber              
 \end{eqnarray}
 
 where $ N $ can be an arbitrary element that satisfies the condition that the matrix $ V $ is unitary. The operator $W$ is a 4$\times$4 sparse matrix $[W_1~\textbf{0}; \textbf{0}~W_2]$ where $\textbf{0}$ is a 2$\times$2 all-zero matrix. $W_1$ and $W_2$ can be illustrated as

  \begin{eqnarray}          
W_1&=&\left(                
  \begin{array}{cccc}   
  \begin{smallmatrix}
   \dfrac{\cos \beta_{1} + \cos \alpha_{1}}{\sqrt{2(1+\cos(\beta_{1}-\alpha_{1}))}} & \dfrac{\cos \beta_{1} - \cos \alpha_{1}}{\sqrt{2(1-\cos(\beta_{1}-\alpha_{1}))}} \\  
    \dfrac{\sin \beta_{1} + \sin \alpha_{1}}{\sqrt{2(1+\cos(\beta_{1}-\alpha_{1}))}} & \dfrac{\sin \beta_{1} - \sin \alpha_{1}}{\sqrt{2(1-\cos(\beta_{1}-\alpha_{1}))}} \\\nonumber 
  \end{smallmatrix}
  \end{array}
\right)    \\
W_2&=&\left(                
  \begin{array}{cccc}   
  \begin{smallmatrix}
  &\dfrac{\cos \beta_{2} + \cos \alpha_{2}}{\sqrt{2(1+\cos(\beta_{2}-\alpha_{2}))}} & \dfrac{\cos \beta_{2} - \cos \alpha_{2}}{\sqrt{2(1-\cos(\beta_{2}-\alpha_{2}))}} \\
   & \dfrac{\sin \beta_{2} + \sin \alpha_{2}}{\sqrt{2(1+\cos(\beta_{2}-\alpha_{2}))}} & \dfrac{\sin \beta_{2} - \sin \alpha_{2}}{\sqrt{2(1-\cos(\beta_{2}-\alpha_{2}))}}\\\nonumber
  \end{smallmatrix}
  \end{array}
\right)  
 \end{eqnarray}

\begin{figure*}
\centering
\includegraphics[width=0.95\textwidth]{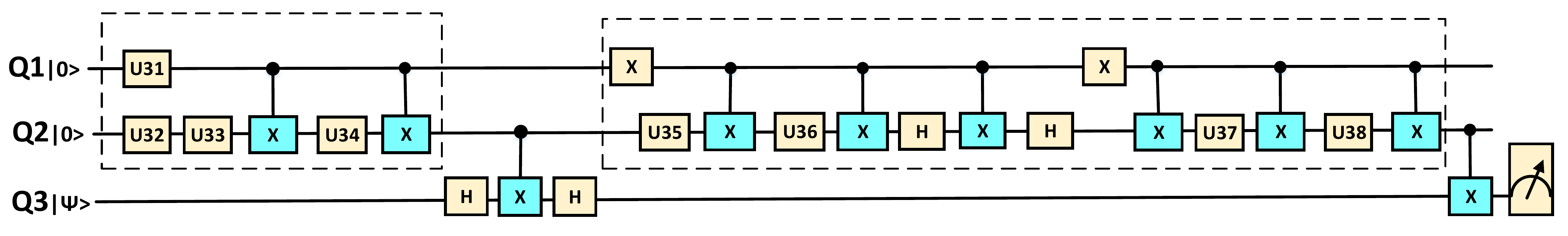}
\caption{  Digital quantum simulation circuit of the universal quantum channel in IBM cloud.  $ Q{1}  $ and $ Q{2} $ form the auxiliary system, and $ Q{3} $ is work qubit. The circuit is simplifyed  with only controlled NOT gates required. Furtherly,  $V$ and $W$ are decomposed into a combination of single qubit operations and controlled NOT gates as dashed box labeled. This  is the final expermental circuit implemented on IBM  qauntum computer. $ U3i $,  $i=[1,2,......,8]  $ is unitary rotation operation $R_{y}(\theta_{i})  $ depend on the concrete elements of $V$ and $W$. } \label{fab}
\end{figure*}
$ K_{j} $ can be expressed by duality gate $ L_{j}=\sum_i W_{ji} V_{i0} U_i, j={0,1,2,3}  $  and a basis changing operation:
 \begin{eqnarray}
  K_{0}&=&(\sum_i W_{0i} V_{i0} U_i) I\nonumber,\quad
  K_{1}=(\sum_i W_{1i} V_{i0} U_i )X\nonumber\\
  K_{2}&=& (\sum_i W_{2i} V_{i0} U_i )I\nonumber,\quad
  K_{3}= (\sum_i W_{3i} V_{i0} U_i)X.\\
  \end{eqnarray}

 Measuring  the final wave functions when the qudit is in state $|j \rangle $ by placing four detectors. The whole process can be denoted as:
\begin{eqnarray}
|00\rangle |\Psi\rangle & \rightarrow & E_{0}|00\rangle |\Psi\rangle +E_{1}|01\rangle |\Psi\rangle   \\\nonumber
&+& E_{2}|10\rangle |\Psi\rangle +  E_{3}|11\rangle |\Psi\rangle. \\\nonumber \label{eq3}
\end{eqnarray}
 We  readout four outputs with the  auxiliary system in state $|00\rangle$,  $|01\rangle$, $|10\rangle$ and  $|11\rangle$ respectively  and finally realized the four Kraus operators  simultaneously. If we only want to obtain the evolution  results of the whole quantum channel, measurement on work qubit is enough. Specially, in the condition that the quantum channel corresponding to two Kraus operators or implementing  with a  classical random number  generator \cite{prlw}, only one auxiliary qubit is required to perform the algorithm.

\section{Experimental results from the IBM quantum computer }
\subsection{Realisation  of universal qubit channel}

\begin{figure}
\centering
\includegraphics[width=0.42\textwidth]{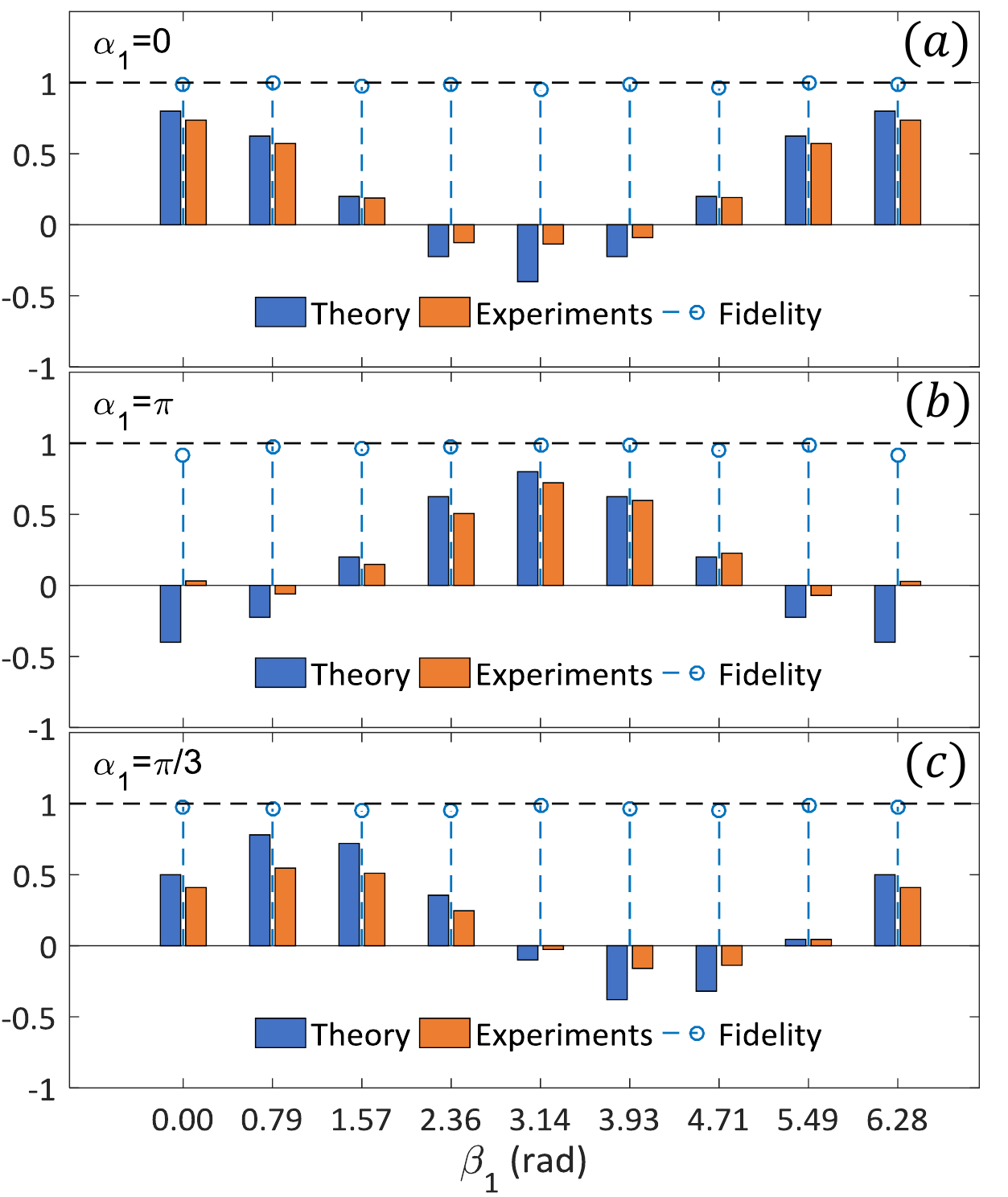}
\caption{ Experimental results from the IBM quantum computer. The $y$ axes means the expected value $ \langle\sigma_{x} \rangle $ and the fidelity of the final state. In detail, the bars show the comparison between theoretical prediction and experimental results for measuring the expected value $ \langle\sigma_{x} \rangle $. Further, the fidelity between experimental final state $\rho_{exp}$ and theoretical state $\rho_{th}$ is illustrated by the circles.}
\label{IBM}
\end{figure}

Experimentally, we utilize the IBM Quantum Experience project in the cloud,  a universal five-qubit quantum computer based on superconducting transmon qubits which has  been tested in various ways\cite{ibm,ibm2,ibm3,ibm4}, to perform our algorithm and compare the experimental result with the ideal quantum channel. We simplify the quantum circuit  with a combination of single qubit gates and controlled NOT gates to carry out the experiment using three superconducting transmon qubits, as shown in Fig. \ref{fab}. We make a measurement on the work  qubit $ Q3 $ to obtain the evolution result after the whole quantum channel.  The experimental fidelity is above 98.5\% in all the following experiments.
 
 \begin{figure}
\centering
\includegraphics[width=0.42\textwidth]{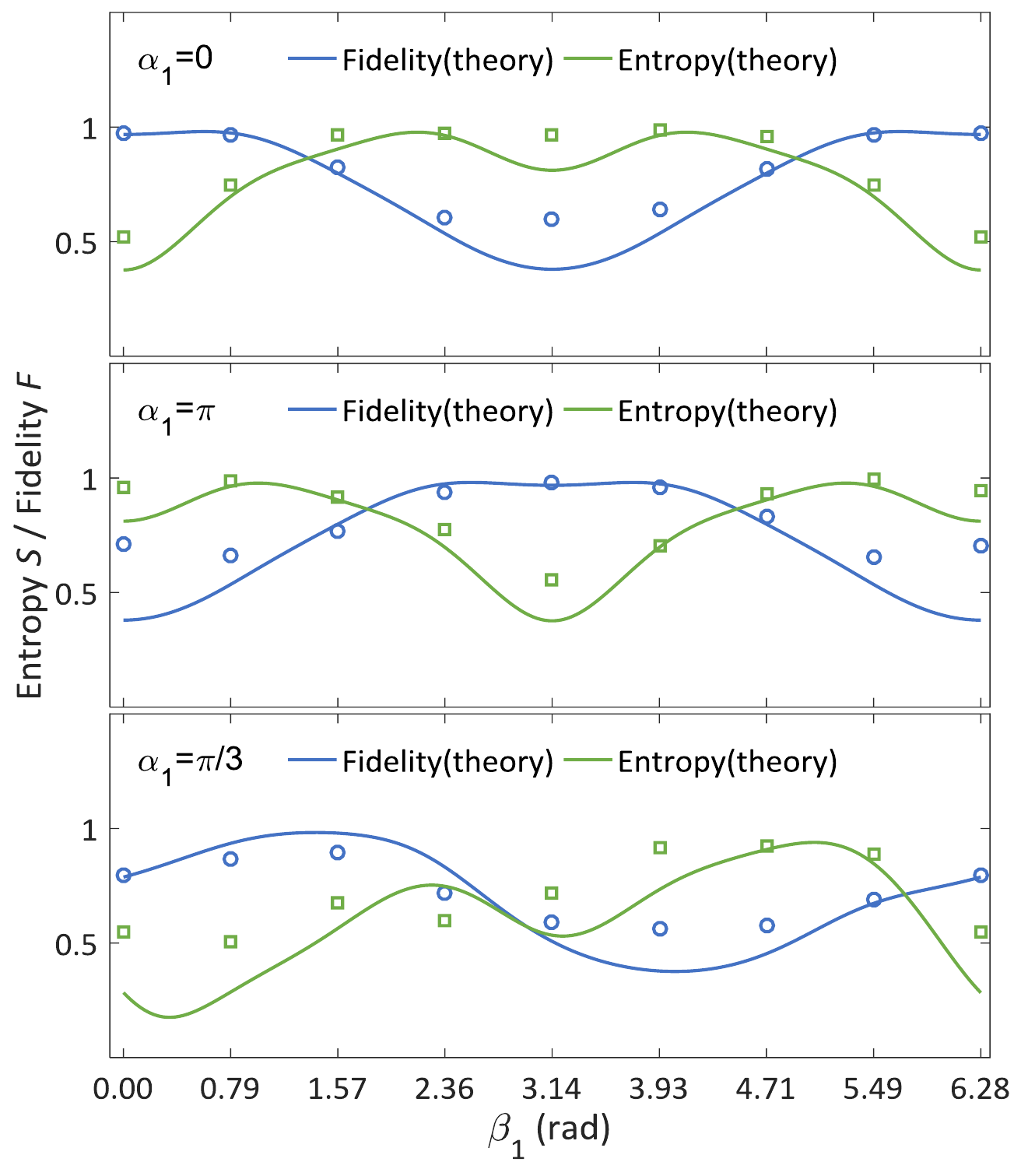}
\caption{ Experimental results from the IBM quantum computer on the analysis for some properties of quantum channel. In detail, the blue lines illustrate the fidelity between the final state and initial state from theoretical preparation. The cyan lines mean the theoretical von Neumann entropy of the final state. The experimental  results are shown by the  corresponding square and circle.} \label{IBM_s}
\end{figure}

In order to  show the feasibility of our algorithm, we totally carried out three classes by changing different parameters in equation (\ref{kkkk}) and  simulate the single  qubit quantum channel with the assist of our algorithm for the initial state $\frac{1}{\sqrt{2}} (|0\rangle+|1\rangle)$. For each class, we merely change parameter $\beta_1$ from 0 to $2\pi$ with the increment $\pi/4$. The detailed setting of the rest parameters: ($a$) $P=0.6,\alpha_1=0, \alpha_2=\pi/2$ and $\beta_2=\pi/6$;   ($b$) $P=0.6,\alpha_1=\pi, \alpha_2=\pi/2$ and $\beta_2=\pi/6$; ($c$) $P=0.6,\alpha_1=\pi/3, \alpha_2=\pi/2$ and $\beta_2=\pi/6$. Each experiment run 8192 shots which means  repeating 8192 times  to decrease statistical errors. In the end of circuit,  the density matrix $\rho$ which reflects the dynamics of the single qubit quantum channel is reconstructed by measuring the expected value of different Pauli operators. Meanwhile, the expected value $ \langle\sigma_{x} \rangle $ and the fidelity of the reconstructed density matrix are presented in Fig. \ref{IBM}. Some properties of single-qubit quantum channel, such as entanglement fidelity and von Neumann entropy , are further analysed by computing the fidelity of the final density matrix with the prepared initial state and the density entropy of final state, whose results are illustrated in Fig . \ref{IBM_s}. Entanglement fidelity  which characterizes how much a system  is modified by the action of a channel,  is use to study the effectiveness of schemes for sending information through a noisy quantum channel  \cite{RMP,entropy}. Entanglement with  environment can be characterized using the entanglement
entropy,
\begin{equation}
S(\rho_{w})=-Tr(\rho_{w}log_{2}(\rho_{w}))
\end{equation}
where $ \rho_{w} $ is the density matrix of work qubit. The calculation of  entropy  is critical  in determining the channels efficiency  in quantum communication and channels capacity \cite{entropy}.

 \begin{figure}
\centering
\includegraphics[width=0.48\textwidth]{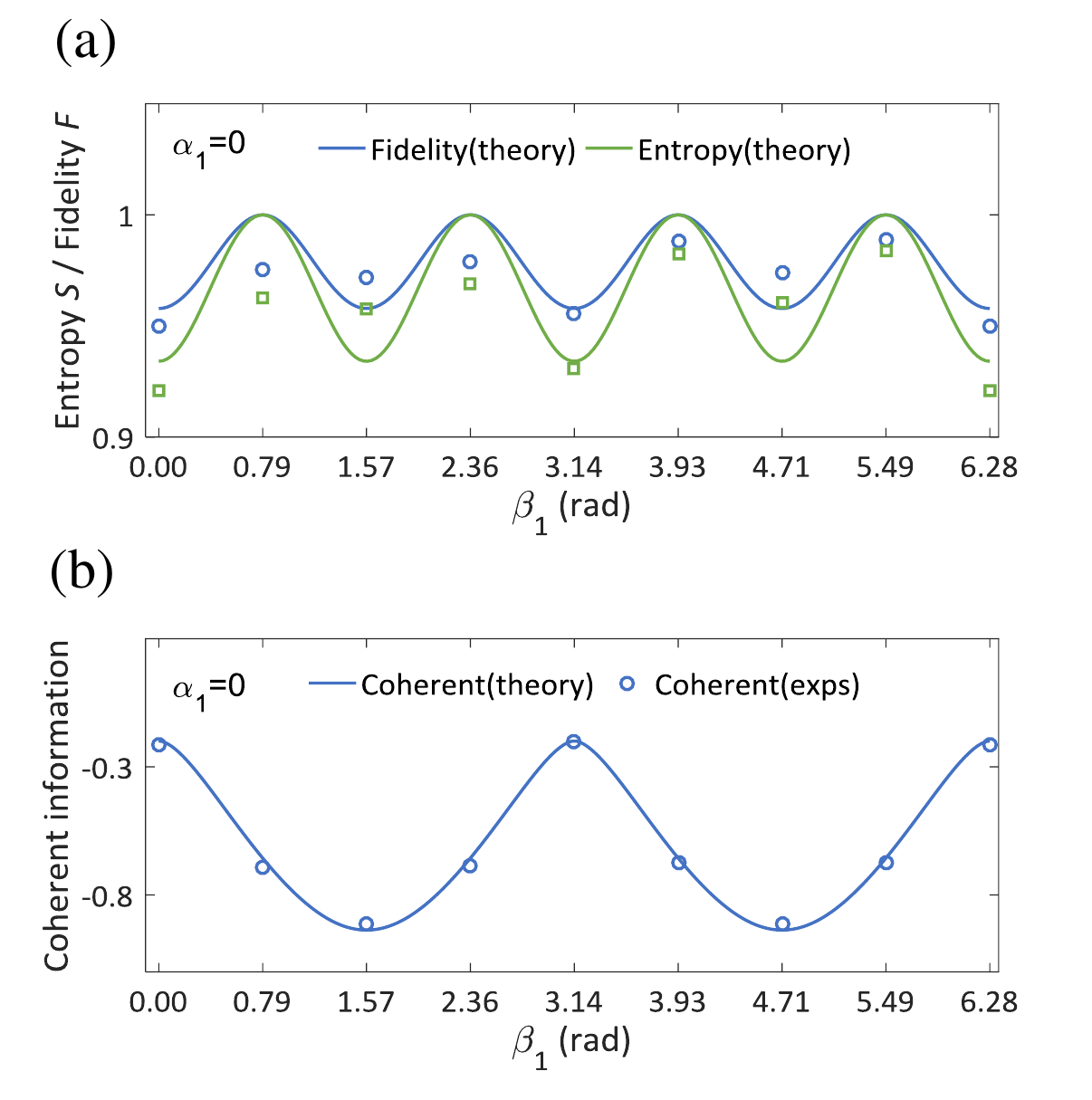}
\caption{Behaviors of work qubit with a mixed initial state through quantum channel. (a) The blue lines and the cyan lines show the theoretical fidelity and von Neumann entropy respectively. (b) The channel coherent information is illustrated by the blue lines. Experimental  results are shown by the  corresponding square and circle with angle $\beta_1$ advances from $ 0 $  to $2\pi$ with the increment $\pi/4$. The rest parameters are fixed as: $P=0.6,\alpha_1=0, \alpha_2=\pi/2$ and $\beta_2=\pi/6$.} \label{IBM_m}
\end{figure}

\subsection{Quantum  channel capacity }

To discuss the  quantum channel more  comprehensively, we show the results of quantum channel on a mixed state in Fig .\ref{IBM_m}. We prepare the mixed 
initial state $\rho_{W}=\frac{1}{\sqrt{2}} (|0\rangle \langle 0| +|1\rangle\langle 1|)  $ which is reduced density operator of a pure state $|\psi\rangle_{WR}=\frac{1}{\sqrt{2}} |00\rangle  + \frac{1}{\sqrt{2}} |11\rangle $ of  a larger system $ WR $ via a purification. In this case, the whole system is consisted of four qubits.

The coherent information roughly measures how much more information work system holds
than environment which is analogous to  mutual information in classical information theory. It is defined as
\begin{equation}
I(\rho,\Phi )=: S[\Phi(\rho)]-S(\rho ;\Phi) ,
\end{equation}
where $S(\rho; \Phi):=  S\left[ (\Phi\otimes {\rm I})(|\psi_\rho\rangle_{WR}\langle\psi_\rho|)\right] $.
The coherent information of the work system after different quantum channels are given in Fig . \ref{IBM_m}.

Furthermore, we can calculate  the quantum channel capacity by a maximization  of the coherent information over all input state \cite{Capacity}
\begin{equation}
C=Max_{\rho} I(\rho,\Phi )
\end{equation}
The quantum capacity of quantum channels is  to quantity the quantum information  can be transmitted coherently through a channel and is a critical factor  in quantum communications \cite{SCI}.
Thus,  we analyse the capacity of some important channels: amplitude damping channel, phase damping channel, and quantum channels corresponding to unital maps.

 \begin{figure}
\centering
\includegraphics[width=0.47\textwidth]{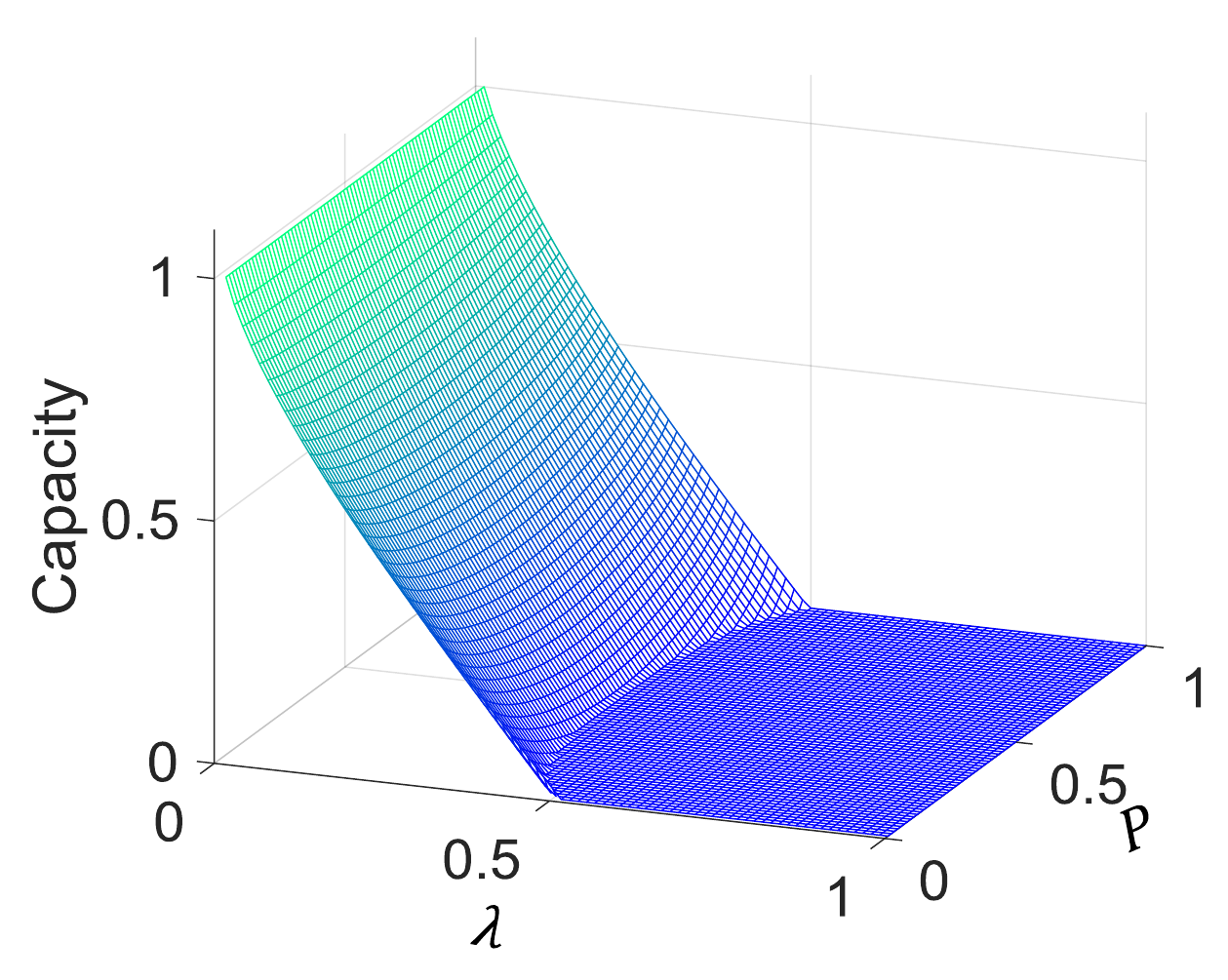}
\caption{Quantum capacity of generalized amplitude damping quantum channels.$ \lambda $ and $ P$ is the  damping rate and  temperature of the environment respectively. } \label{AD}
\end{figure}

A generalized amplitude damping (AD) channel describes the effect of dissipation to an environment at finite temperature \cite{N}. The AD channel which is  widely inhered in various quantum systems  is a critical factor effecting the precision of quantum computation and capacity of quantum communication. For single qubits, it squashes the Bloch sphere towards a
given state, denoted as a $\Phi: \mathcal{M}_2 \rightarrow \mathcal{M}_2$ map.    Setting $ \cos (2 \beta_{1})=1 $ and $ \cos (2 \alpha_{1})= 1-2\lambda$, equation (\ref{kkkk})  defines the amplitude damping channel with damping rate $ \lambda $. The damping parameter $ \lambda $ describes the rate of dissipation and parameter $P$  represents the temperature of the environment.

  We numerically calculate the capacity of generalized amplitude damping channel with different $ \lambda $ and $ P$. As shown in the picture, with the increasing of    dissipation  rate in different temperatures of the environment, the quantum capacity is decreased  from maximum values to  zero. The physical picture is that when the  dissipation  rate at a finite environmental temperature is large enough, the  quantum data  can be protected is decreased to zero. It  can be used as a standard on measuring the ability of quantum systems  that protecting quantum information from the noise environment. The calculation gives a metric on  how reliable  and efficient  of a quantum system to process information undergoes  a amplitude damping channel.

\begin{figure}
\centering
\includegraphics[width=0.47\textwidth]{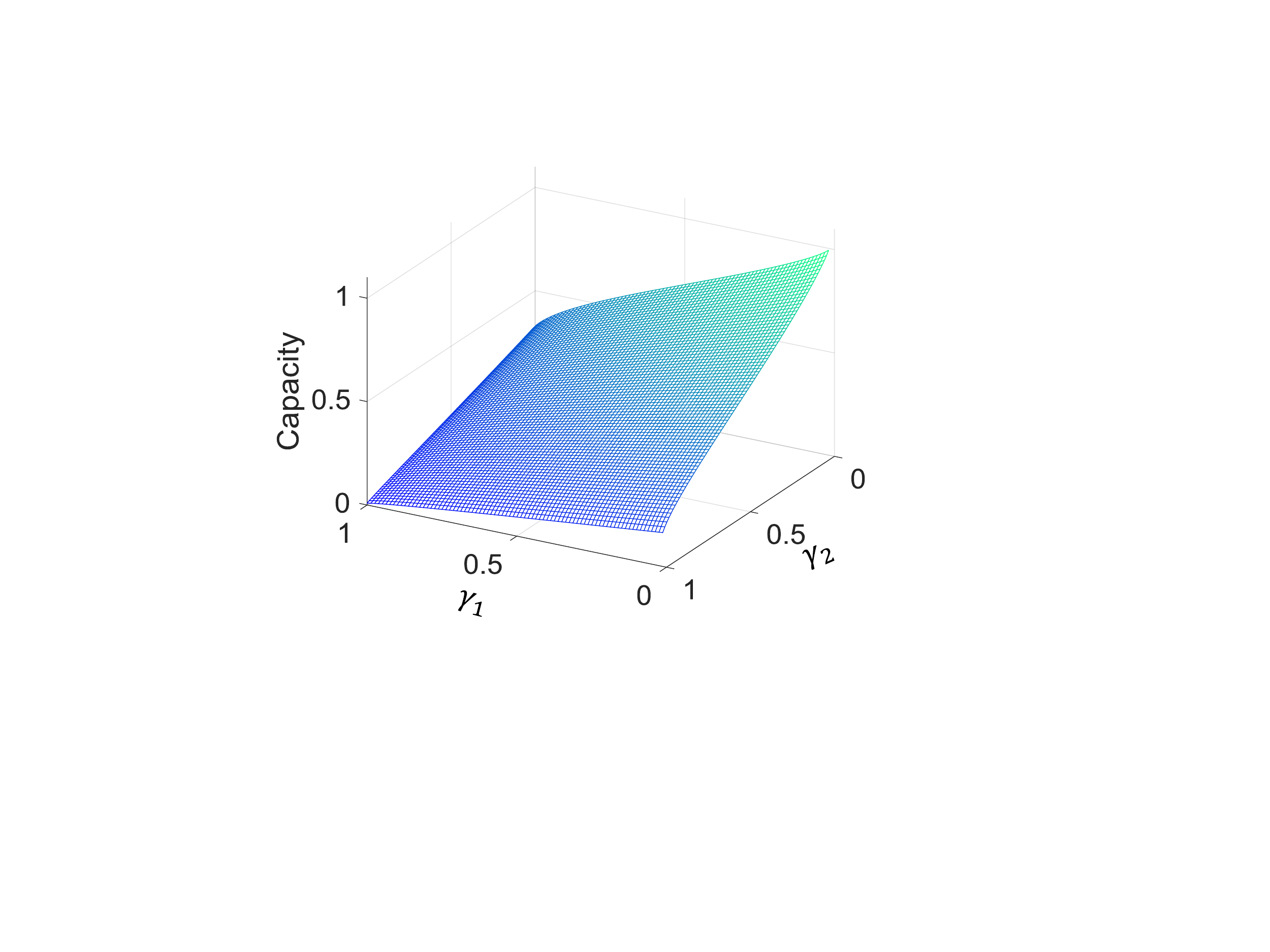}
\caption{Behaviours of  phase damping quantum channel capacity. $ \gamma_{1} $  and  $\gamma_{2}  $  are the two different strengths of the PD channel. } \label{PD}
\end{figure}

A noise channel which describe the loss of quantum information without loss of energy is phase damping (PD) channel\cite{N}. It is quantum mechanical uniquely and is one of the most subtle  process in  quantum computation and quantum information, which has drawn an immense amount of study and speculation. It is regarded as a general environment  effect leading to our world to be so classical by decreasing and even eliminating  coherent information. The phase damping  qubit channel squashes the Bloch sphere towards $ z $ axis and can be expressed as following  
\begin{equation}\label{pdee} 
K_{0}=\sqrt{P}\left(                 
  \begin{array}{cccc}   
    1 & 0 \\  
    0 & \sqrt{1-\gamma_{1}}  \\  
\end{array}
\right),                 
K_{1}=\sqrt{P}\left(                 
  \begin{array}{cccc}   
    0 &0 \\  
   0& \sqrt{\gamma_{1}} \\  
\end{array}
\right), \\
\end{equation}\nonumber
\begin{equation}
K_{2}=\sqrt{1-P}\left(                 
  \begin{array}{cccc}   
   1 & 0 \\  
    0 & \sqrt{1-\gamma_{2}}  \\  
\end{array}
\right),                 
K_{3}=\sqrt{1-P}\left(                 
  \begin{array}{cccc}   
     0 &0 \\  
   0& \sqrt{\gamma_{2}}  
\end{array}
\right). \\\nonumber 
\end{equation}
Where the parameter $ \gamma_{1} $  and  $\gamma_{2}  $  can be interpreted  as the strength of the PD channel,  corresponding to coherence time $T_{1}  $  and  $T_{2}  $ respectively. It is more general than the PD channel expressed by two Kraus operators which can be obtained in the condition $ P=1 $ from Eq. (\ref{pdee}). Considering the unitary freedom of quantum operation, the phase damping channel is equal to a 
recombination of unital channels $ \sigma_{z} $ and $ I $ which can be realised by the universal quantum channel directly. With the coherent strength  or coherent time increasing, less and less quantum information can be protected. In the real word, generally, the quantum system mainly undergoing AD and PD channels because of the inevitable  coupling with environment. The capacity calculations of AD and PD channels are important to  quantify the efficiency to transport quantum information in open quantum system under noise  environment.

\begin{figure}
\centering
\includegraphics[width=0.47\textwidth]{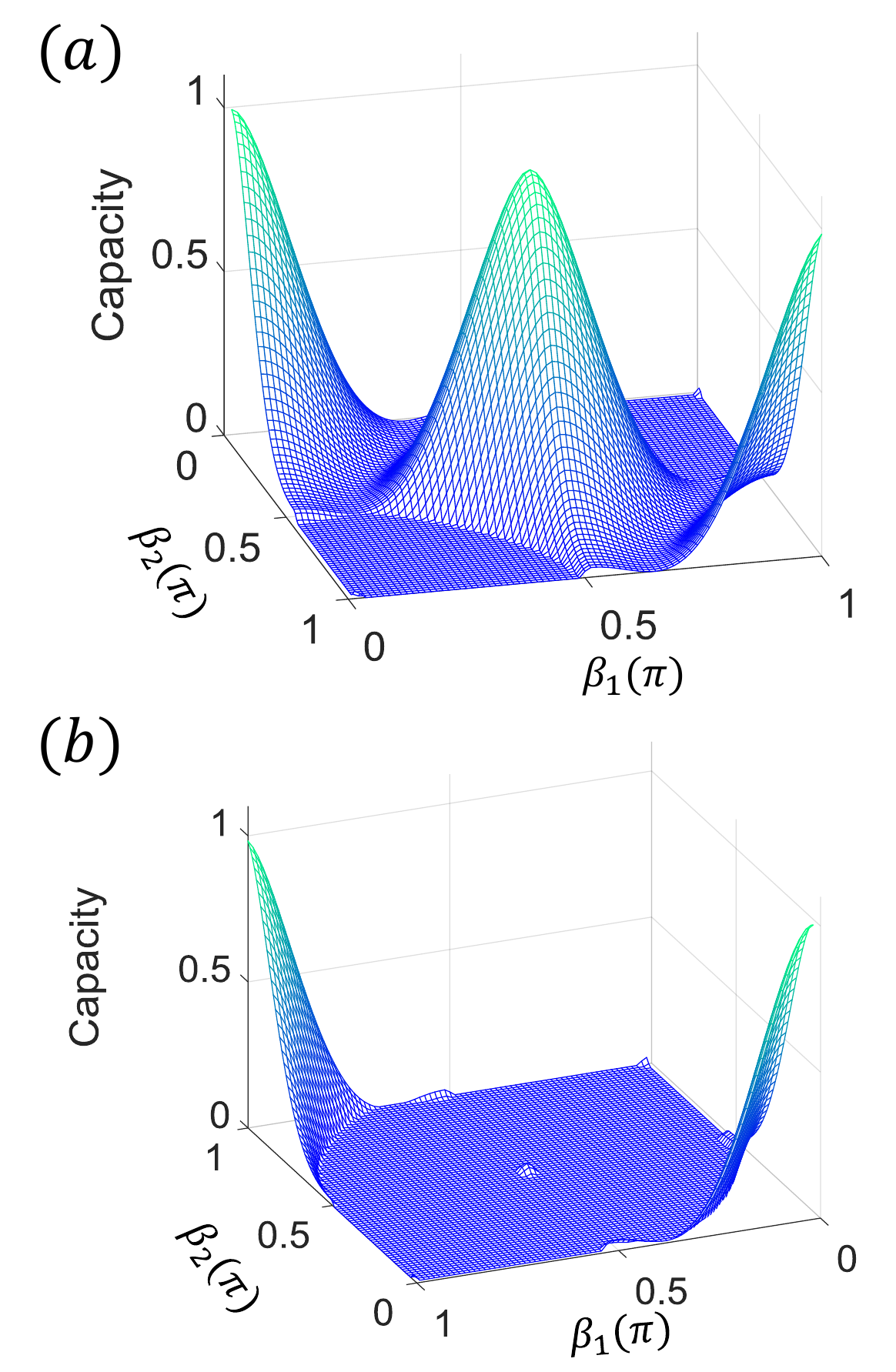}
\caption{Capacity of   quantum channels corresponding to unital maps. Setting  $ P=0.6 $, $ \beta_{1} $  and $ \beta_{2} $ increase from $ 0 $ to $ \pi $ by step $ \pi/80 $.} \label{UM}
\end{figure}

For $\sin\beta_{1}= \pm \sin\alpha_{1} $ and  $\sin\beta_{2}= \pm \sin\alpha_{2} $, one gets unital channels from Eq. (\ref{kkkk}). Specifically, the depolarizing  channel including entanglement breaking (EB) channels is an important type unital channel which transforms  the  initial state towards the centre of the Bloch sphere.
The capacity of all unital qubit channels is illustrated  in Fig. \ref{UM}. In the case $\sin\beta_{1}=  \sin\alpha_{1} $ and  $\sin\beta_{2}= \sin\alpha_{2} $, the capacity reaches maximum to $ 1 $ when $ \beta_{1}= \beta_{2}={0 , \pi/2 , \pi}$ as shown in Fig. \ref{UM}(a). In Fig. \ref{UM}(b),  $\sin\beta_{1}=  \sin\alpha_{1} $ and  $\sin\beta_{2}= -\sin\alpha_{2} $, there are two maximum values  when $ \beta_{1}= \beta_{2}={0 , \pi}$.

\begin{figure}
\centering
\includegraphics[width=0.47\textwidth]{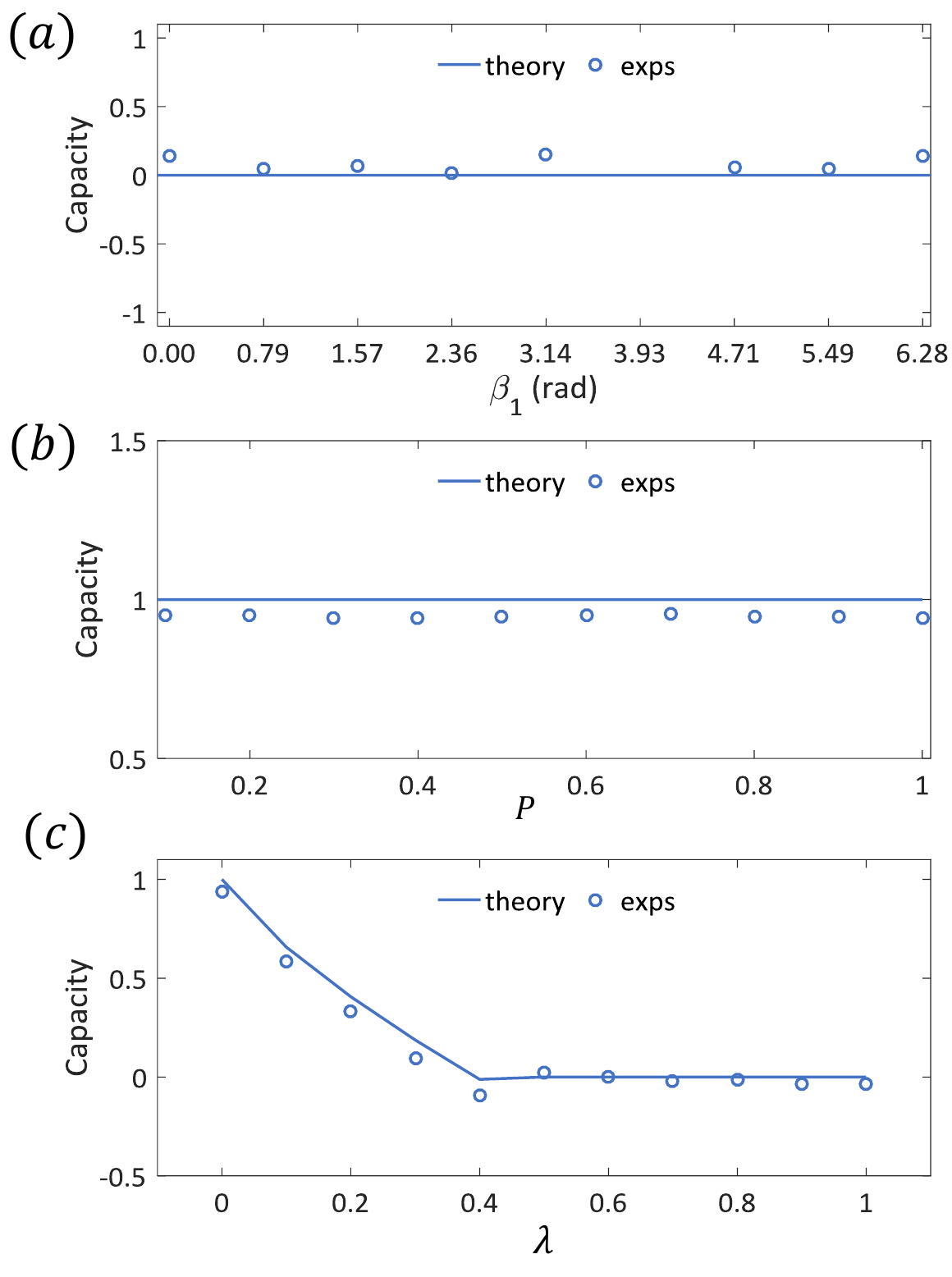}
\caption{Three types of quantum  channel capacity. ($a$) Capacity zero channels with  parameter $\beta_1$ changing from 0 to $2\pi$ with the increment $\pi/4$. The rest parameters are fixed as $P=0.6,\alpha_1=0, \alpha_2=\pi/2$ and $\beta_2=\pi/6$.   ($b$) Capacity maximum channels with $ P $  increasing from $ 0.1 $ to $ 1$ by ten steps $\alpha_1=\beta_1=\alpha_2=\beta_2=\pi/2$.  ($c$) Amplitude damping channels with dissipation rate $ \lambda $ increasing from $ 0 $ to $ 1 $ with the increment $0.1$, $P=0.6,\beta_1=\alpha_2=0, \alpha_1= \beta_1 =\arccos (\sqrt{1-\lambda}).$ The theoretical capacity is illustrated by the blue lines  and experimental  results are shown by the  corresponding circles. } \label{EC}
\end{figure}

We experimentally reveal the behaviour of three types of quantum  channel capacity: capacity zero channel, amplitude damping channel and capacity maximum channel.
In fig. \ref{EC}(a), a group of Zero-Capacity Channels are shown. With the input state $\frac{1}{\sqrt{2}} (|0\rangle+|1\rangle) $,   the coherent information  reaches its largest value zero and turn to be the capacity.
 When it comes to the  channel  with maximum capacity, we obtain the capacity by calculate the coherent information with maximum mixed  state  as initial state in fig. \ref{EC}(b).In fig. \ref{EC}(c), fixing $P=0.6  $, we show  the capacity of amplitude damping channel within the different dissipation rate $ \lambda $. When $ \lambda < 0.4$, the capacity is corresponding to  the  coherent information with the maximum  mixed  state  as input state. When $ \lambda \geq 0.4$, the capacity is corresponding to  the  coherent information with the pure state  as input state.

Based on the above results, it is believed that the experimental results agree well with the theoretical predictions within a certain errors. Considering that we have repeated the experiment enough times, the statistical errors are reduced. The systematic errors which is mainly contributed by single gate  and controlled NOT gates errors  is the most important factor leading to the discrepancy with the ideal results.

\section{Discussion}
In conclusion, we present a new method to simulate the dynamics of open quantum system by performing the Kraus operators in a linear combination of unitary operators form. We have experimentally shown how to realize an arbitrary single qubit channel which can be regarded as a primitive for simulating open quantum system dynamics. Our algorithm only requires single direction controlled operation from ancillary system to work qubit and realizes four Kraus operators which correspond to a universal qubit channel simultaneously.  
Additionally,  the ability to obtain the density information of work qubit under  any single Kraus operator makes our algorithm flexible  and feasible.  This method is general and scalable  to construct algorithm to simulate the open quantum system  dynamics in higher dimension.
In our algorithm, the dimension of the ancillary system is determined by the maximum value of number of unitary operators and number of Kraus operators. The maximum value of both are equal to $d_s^2$, where $d_s$ is the dimension of the work-qubit system.
Consider the fact that any matrix can be written as a  linear combination of four unitary matrices {\cite{sup}}, the Kraus operator represented  quantum channel can  be decomposed into a linear combination of four unitary operators in arbitrary dimension. In detail, if the Kraus operators is less than $ 4 $, only $ 2 $ ancillary qubits are required to perform a quantum channel in any dimension. In the condition that a channel has the number of kraus operators more than  $ 4 $, 
we can realize arbitrary dimension quantum channel  in the form of convex combination of  Kraus operators with the assistance of a number of ancillary qubits that grows  logarithmically in the number of Kraus operators.
Thus, only in the worst case that the number of  Kraus operators is equal to $d_s^2$, our method require the same ancillary resource as the standard dilation of a channel.

 In our approach, the efficiency is mainly  reflected in the gate complexity. According to Ref. \cite{N, complex1, complex2}, a $M$-qubit arbitrary unitary operation can be implemented using a circuit containing $O(M^3 4^M)$ single qubits and controlled NOT gates.  Now, considering a $n$-qubit original system with  a $2n$-qubit environment, the gate complexity of standard Stinespring dilation method is $O(27n^3 4^{3n})$. In our method, the  unitary operations $V$ and $W$ performed on the ancillary system can be decomposed  into $O(8n^3 4^{2n})$single qubits and controlled NOT gates. The controlled operations between $V$ and $W$ can be decomposed into $n4^{2n}$ single qubit gates and $4^{2n}$ CNOT gates. Therefore, the total gate complexity of our method is $O(8n^3 4^{2n}+n4^{2n}+4^{2n})$. For the large system, it is clearly showed that the improvement of performance is significant compared with Stinespring dilation.

Furthermore, we explore the universal qubit quantum channel properties and calculate the quantum capacity of different channels.  The analysis  of amplitude damping, phase damping, unital maps channels capacity provides  a potential application in quantum communication and information.

The authors would like to thank Bei Zeng for helpful discussions. This work was supported by the National Basic Research Program of China (2015CB921002), the National
Natural Science Foundation of China Grant Nos. (11175094, 91221205). Wei is supported by the Fund of Key Laboratory (9140C75010215ZK65001).

\clearpage

\onecolumngrid

\section*{Supplemental Material for \\ ``Efficient universal quantum channel simulation in IBM's cloud quantum computer''}

In this supplemental material, we provide some theoretical and experimental details of the employed setup and techniques. 
\\

\section*{ THE THEORY OF IMPLEMENTING NON-UNITARY OPERATORS}
The whole process of realising non-unitary operator is shown as following:
First, performing the unitary operator $V$ on the auxiliary system to construct a superposition state. Secondly, we implement the
auxiliary system controlled operations $U_{0}\otimes |0\rangle \langle 0| , U_{1}\otimes |1\rangle \langle 1| ,\ldots , U_{d-1}\otimes |d-1\rangle \langle d-1|$ on the work system. Then, performing the unitary operation $W$ on the auxiliary system. Usually, at this stage, we have realized all the transformations in  duality quantum computing. Finally, we readout the results by observing in the subspace of  ancillary system. There are at most $d$ outputs in one process by measuring all the eigenvalue states of the ancillary system.

The whole process can be expressed as
\begin{eqnarray}
|\Psi\rangle|0\rangle &\rightarrow & \sum_{i=0}^{d-1}V_{i0}|\Psi\rangle|i\rangle\nonumber\\		
&\rightarrow &\sum_{i=0}^{d-1}V_{i0} U_i|\Psi\rangle |i\rangle\nonumber\\
&\rightarrow &\sum_i V_{i0} U_i|\Psi\rangle W|i\rangle\nonumber\\
&=&\sum_i\sum_j W_{ji} V_{i0} U_i|\Psi\rangle |j\rangle \nonumber\\
&= & \sum_j L_{j}|\Psi\rangle |j\rangle,
\nonumber
\end{eqnarray}
where $ L_{j}= \sum_i W_{ji} V_{i0} U_i $  is denoted as the duality quantum gate and $W_{ji} V_{i0}$ is the complex coefficient which satisfies $\sum_{i=0}^{d-1}| W_{ji} V_{i0}|\le 1$. The duality quantum gate composed by a linear combination of unitary operations is the key to realize Kraus operator.

\section*{ MATRIX DECOMPOSITION}
Any matrix can be written as a  linear combination of four unitary matrices. The  proof is given as following.
Define $ A $ as a normalize complex matrix satisfying $ \Vert A\Vert\leqslant 1 $. Then
\begin{equation}
A=B+iC
\nonumber
\end{equation}
where $B  $ and $ C$ are selfadjoint and given by
\begin{eqnarray}
B=\frac{1}{2}(A+A^{\dagger}), C=\frac{1}{2i}(A+A^{\dagger}).
\nonumber
\end{eqnarray}
The two selfadjoint operators  satisfy $\parallel B \parallel\leqslant 1 $ and $\parallel C \parallel\leqslant 1 $, which means that their eigenvalues in the range $[-1,1]$. Then , $B  $ and $ C $ can be decomposed as
\begin{equation}
B=\frac{1}{2}(U_{1}+U_{2}), C=\frac{1}{2}(U_{3}+U_{4}).
\nonumber
\end{equation}

where$ U_{1},U_{2},U_{3},U_{4} $  are unitary and given by
\begin{eqnarray}
U_{1}=B+i\sqrt{I-B^{2}},U_{2}=B-i\sqrt{I-B^{2}}\nonumber\\
U_{3}=C+i\sqrt{I-C^{2}},U_{4}=C-i\sqrt{I-C^{2}},
\nonumber
\end{eqnarray}\

Then $ A=\frac{1}{2}(U_{1}+U_{2})+\frac{i}{2}(U_{3}+U_{4}) $ is a linear combination of four unitary matrices.

\section*{CHIP ARCHITECTURE AND MEASUREMENT SETUP} 
 The present experiment is performed using three and four superconducting
transmon qubits in IBM quantum computer. The superconducting circuits are transmon qubits wuth resonance frequencies between $5.06$ and $5.3\:$GHz  connected by Josephson junctions and shunt capacitors that provide superpositions of charge states. The  connections among individual qubits and the classical control system are realised  by waveguide resonators.  The operations and measurement are achieved by applying tailored microwave signals to this system and measuring the response. Qubits are resolved in the frequency domain during addressing and readout. In the Quantum Experience hardware, it provides four two-qubit interactions. Only the qubits which have interactions, The CNOT gates are allowable. Single-qubit readout fidelities are about $\sim96\%$  and typical gate fidelities are $99.7\%$ and $96.5\%$ for single qubit gate and the CNOT gates respectively. The pulse time to perform typical single gates and CNOT gates are $130\:$ns  and $250-450\:$ns respectively. The coherence times of two channels  both amplitude damping ($T_1$) and spin dephasing ($T_2$) are shown as following.
\begin{tabbing}
\hspace{1.5cm}\=\hspace{1.5cm}\=\hspace{1.5cm}\=\hspace{1.5cm}\=\hspace{1.5cm}\=\kill
 Item   \>  Q1 \>  Q2 \>  Q3 \>  Q4 \> Q5 \\ 
 T1($ \mu s $) \>  58.2 \>  68.1 \>  44.4 \>  48.3 \> 54.1 \\ 
 T2($  \mu s$)\>   52.6 \>  40.7 \>  71.7\>  57.5 \> 88.7
\end{tabbing}

\end{document}